# Anomalous Photon-induced Near-field Electron Microscopy

Yiming Pan[1,3*], Bin Zhang[2*] and Avraham Gover[3]

1. Department of Physics of Complex Systems, Weizmann Institute of Science, Rehovot 76100, ISRAEL
2. National Laboratory of Solid State Microstructures and School of Physics, Nanjing University, Nanjing 210093, CHINA
3. Department of Electrical Engineering Physical Electronics, Tel Aviv University, Ramat Aviv 69978, ISRAEL

**Abstract**

We reveal the classical and quantum regimes of free electron interaction with radiation, common to the general variety of radiation sources (e.g. FEL, Smith-Purcell), Dielectric Laser Accelerator (DLA) and Photo-Induced Near-Field Electron Microscopy (PINEM). Modelling the electron with initial conditions of a coherent quantum electron wavepacket (QEW), its topology in phase-space uniquely defines a universal distinction of three interaction regimes (and their particle-wave duality transition): point-particle-like acceleration, quantum wavefunction (PINEM), and a newly reported regime of anomalous PINEM (APINEM). The quantum interference beat of APINEM is capable of improving the spectral resolution of post-selective electron microscopy, and the particle-wave duality transition reveals the history-dependent nature of quantum electron interaction with light.

Here we address a large class of light-matter interaction schemes and devices, in which free electrons are stimulated to emit/absorb light quanta (i.e., photons) when interacting with a coherent radiation field (laser beam). This class of radiative interactions includes numerous schemes of radiation sources and laser accelerators, such as Free Electron Laser (FEL), Cherenkov radiation [1-3], Smith-Purcell Radiators (SPR), Cherenkov radiation, Transition Radiation (TR) [6] and Dielectric Laser Acceleration (DLA) [1-8]. On the other hand, this class also includes the advanced ultrafast electron microscopy schemes of Photon-Induced Near-field Electron Microscopy (PINEM) [9-11] and coherent manipulation of quantum wavefunction with light [11-13].

The PINEM-kind schemes are based on a quantum process of multiphoton emission and absorption of light quanta ($\hbar\omega$) that takes place simultaneously when an electron wavefunction passes through the near field or the confined plasmonic excitation field of a nanostructure, nano-tip or foil, illuminated by an ultrafast laser beam pulse [4-5, 11-13, 18-19]. This can be described as a stimulated TR process, in which the monoenergetic spectrum of the passing-by electron develops discrete symmetric photon sidebands, spaced $\hbar\omega$ apart ($\omega$ is the frequency of the incident laser field) due to discrete energy quanta emission and absorption from the incident radiation field. In order to distinguish the energy sidebands in the PINEM spectrum in the quantum mechanical operating regime of electron-photon interaction, one must require that the photon energy spacing of the sidebands exceeds the energy spread $\Delta E$ of the beam [9, 14]:

$$\Delta E < \hbar\omega, \text{ or } \Delta p < \hbar\omega/v_0, \qquad (1)$$

This condition is equivalent to the "large recoil condition" for quantum FEL [14, 20] where $\hbar\omega/v_0$ is the emission/absorption electron quantum recoil momentum, and $\Delta p = \Delta E/v_0$ is the energy momentum spread, $v_0$ is the electron group velocity.

In this letter, we refer to the entire class of all of these free electron stimulated radiative interaction schemes (Fig.1a), and report the operating characteristics of a hitherto non-investigated light-matter interaction regime, where the electron is not presented as a plane-wave, as in conventional quantum FEL [20,15] and PINEM [9-12] models, or as a point-particle, as in classical electrodynamics models of FEL [2,3] and accelerators [4,5], but instead as a finite duration and finite energy spread coherent Quantum Electron Wavepacket (QEW) (e.g., a Gaussian envelope wavefunction) satisfying the minimal Heisenberg uncertainty: $\sigma_{E_0}\sigma_{t_0} = \hbar/2$, $(\sigma_{p_0}\sigma_{z_0} = \hbar/2)$, where $\sigma_{E_0}$ ($\sigma_{p_0} = \sigma_{E_0}/v_0$) is the standard deviation of the wavepacket energy (momentum) and $\sigma_{t_0}$ ($\sigma_{z_0} = v_0 \sigma_{t_0}$) is the standard deviation of its duration (spatial size) at the minimal waist point of the Gaussian wavepacket propagation. As the wavepacket drifts in free space, its intrinsic energy spread is retained, but the wavepacket size expands as a function of drift time $t_D$ according to:

$$\sigma_z(t_D) = \sqrt{\sigma_{z_0}^2 + \left(\frac{\lambda_c^*}{4\pi} \frac{ct_D}{\sigma_{z_0}}\right)^2},$$ where we defined $\lambda_c^* = \lambda_c/\gamma^3$ with $\lambda_c = h/mc$ – the Compton wavelength[17].

The free propagation of the QEW leads to spatial stretching in the axial dimension, accompanied by energy (phase) chirping effect. This expansion suggests a different condition for operating in the quantum (PINEM) regime, where the wavepacket, long enough, acts as a plane wave with no phase relation to the light wave - the "long wavepacket" condition[17,20]:

$$2\sigma_t(t_D) > T = 2\pi/\omega, \quad (2\sigma_z(t_D) > \lambda\beta/2\pi). \tag{2}$$

Here $\beta = v_0/c$ and $\lambda$ is wavelength of light. In order to understand the validity ranges of these two different quantum limit conditions (1-2), it is most instructive to present the electron at entrance and after interaction in an energy-time (E-t) phase-space, as shown in Fig.1, or in the corresponding momentum-space (p-z) phase-space. Based on the topology of the electron distribution in this phase-space, we report here distinction between three universal operating regimes, common to all light-electron interaction schemes: Acceleration, PINEM and anomalous PINEM (APINEM). This distinction depends solely on the initial conditions of the wavepacket, and the phase-space presentation demonstrates the transition of these schemes from the point-particle classical regime to the quantum regime. We thus define (here in p-z phase space) the Wigner Distribution (WD) representation of the QEW [15], i.e., $W(z,p,t) = \frac{1}{2\pi} \int \psi^*(p-q/2)\psi(p+q/2) e^{-i(E_{p+q/2}-E_{p-q/2})t/\hbar} e^{iqz/\hbar} dq$, where $\int W(z,p,t) dzdp = h/2$ and $E_p \simeq \varepsilon_0 + v_0(p-p_0) + (p-p_0)^2/2m^*$ is the relativistic energy dispersion, expanded to second-order around the incoming energy $\varepsilon_0$ [17, 20].

*First order perturbation analysis* - In a first-order perturbation solution of Schrodinger equation (see supp. A), the perturbed wavefunction after interaction, is given in momentum space as:

$\psi(p) = \psi^{(0)}(p) + \psi^{(1)}(p)$, where $\psi^{(0)}(p)$ is the initial wavefunction and $\psi^{(1)}(p)$ the scattered component. Thus, after interaction, the laser-induced Wigner function is composed of three terms:

$$W(z,p) = W^{(00)}(z,p) + 2\Re\left(W^{(01)}(z,p)\right) + W^{(11)}(z,p), \qquad (3)$$

where $W^{(00)} = \frac{1}{\aleph}\exp\left(-\frac{(z-pt_D/m^*)^2}{2\sigma_{z_0}^2} - \frac{(p-p_0)^2}{2\sigma_{p_0}^2}\right)$ is the initial Gaussian WD, rescaled by normalization ($1/\aleph$), $W^{(11)}$ is the scattered term, and $2\Re\left(W^{(01)}\right)$ is an interference term between the zero order and the first order scattering term of the wavefunction. Keeping this interference term, which has been neglected in previous quantum analyses, is a pivotal methodological step in this work, since it is essential for the identification of the APINEM effect and the point particle to quantum transition in the QEW regime.

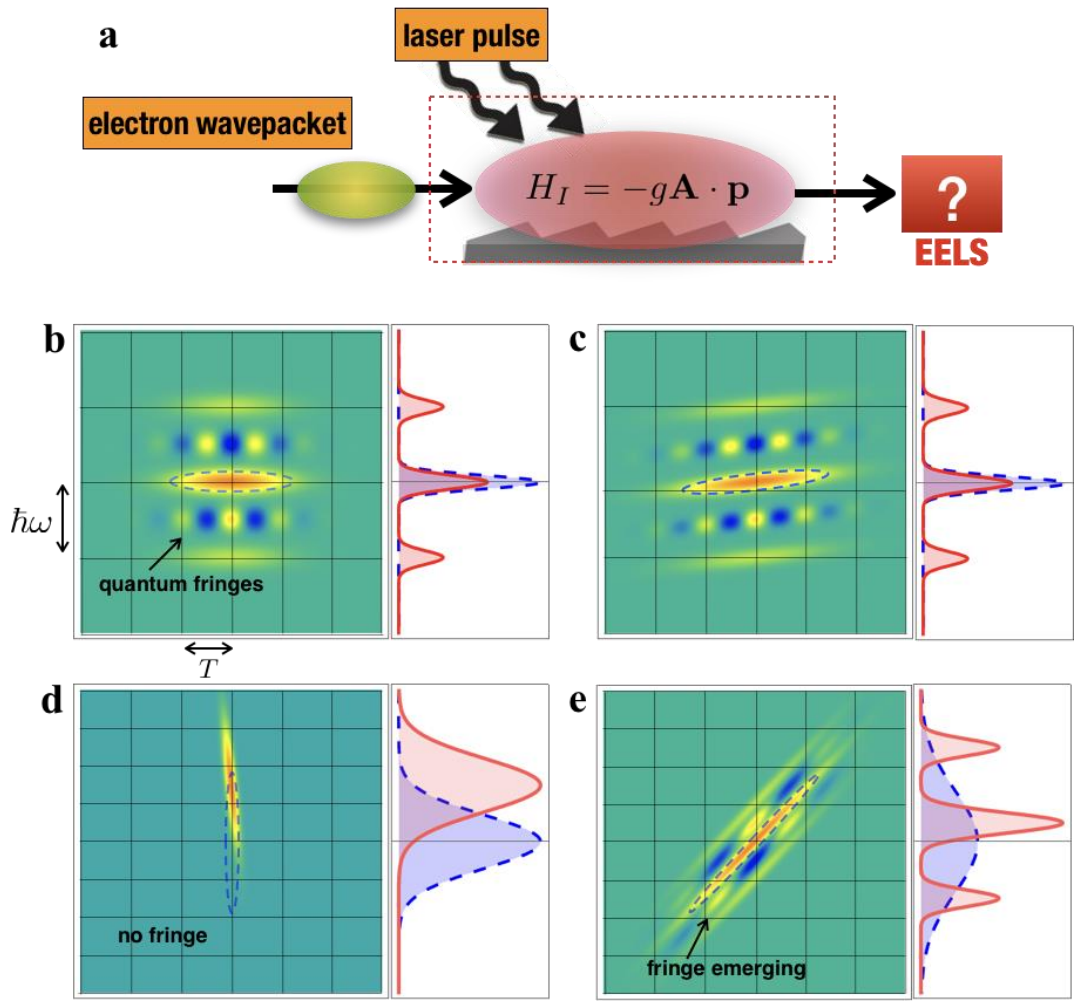

*Fig.1: Illustrations of PINEM, Acceleration and APINEM processes in phase space representation before (broken-line ellipses) and after interaction (positive- red, yellow, negative- blue) and their momentum distributions (EELS spectrum). (a) The universal light-electron interaction scheme. (b-c) PINEM with quantum fringes in-between the photon sidebands, for expanded un-chirped (b) and pre-chirped (c) quantum electron wavepackets. (d) Particle-like acceleration with net momentum shift. (e) APINEM with quantum fringes emerging.*

Fig. 1b-d displays the pre-interaction WD of the QEW (in broken line) and the post-interaction WF (in color code) in (E-t) phase-space, overlayed over a quantization grid of $\hbar\omega \times T = h$, with $T = 2\pi/\omega$ (or $(\hbar\omega/v_0) \times (\beta\lambda) = h$ in p-z phase-space). It reveals the different possible interaction regimes in dependence on the QEW initial condition, based alone on the topology in phase space of the initial electron WF. Note that the Gaussian QEW distribution has a phase area of half Planck constant at the minimal waist point ($2\pi\sigma_{E_0}\sigma_{t_0} = h/2$), and this area stays constant under the horizontal stretching transformation of drift before interaction. For QEW entering the interaction region at its waist, namely as an erect ellipse with no chirp (Fig.1b&d), the phase-space topology leaves only two qualitatively different scenarios. In one case (Fig.1b) the WD extrudes out of the quantization box in the horizontal dimension, satisfying the "long wavepacket" quantum regime condition (2) - $2\sigma_{t_0} > T$, and because $\sigma_{t_0} = \hbar/2\sigma_{E_0}$, it satisfies also the "large recoil" condition (1): $2\sigma_{E_0} < \hbar\omega$, (or $2\sigma_{p_0} < \hbar\omega/v_0$). The other possibility (Fig.1d) is a narrow wavepacket case, $2\sigma_{t_0} < T$ (narrow broken line ellipse), that necessarily corresponds also to extruding out of the quantization box in the vertical dimension: $2\sigma_{E_0} > \hbar\omega$, (or $2\sigma_{p_0} > \hbar\omega/v_0$), namely, violating both kinds of quantum regime conditions (1-2). After interaction, the $\pm\hbar\omega$ vertically shifted energy sidebands do not overlap in case (1b), and their horizontal projection produce the PINEM-kind multi-sidebands energy spectrum as shown aside the WD picture. On the other hand, in case (1d), where the vertical projection corresponds to a point-particle-like wavepacket with distinguishable phase relative to the radiation wave, the horizontal projection of the distribution produces a point-particle-like acceleration spectrum of the classical limit.

If the electron arrives to the interaction region after drift (Fig.1c&e), the stretching transformation of free drift produces a chirped long wavepacket distribution at the entrance to the interaction region (slanted broken line ellipse extruding horizontally out of the quantization box), satisfying in both cases the "long wavepacket" PINEM condition (2). In case (1c) this condition is still kept consistently with the "large recoil" condition, since the momentum $\sigma_{p_0}$ does not change in drift, and consequently the horizontal projection of the post-interaction WD produces the conventional multi-sidebands PINEM spectrum as in (1b).

Of special interest is the case (1e), where the narrow QEW distribution of (1d), stretching due to drift before entering the interaction region, satisfies the long wavepacket condition, but violates the conventional PINEM quantum recoil condition (1). Nevertheless, the horizontal projection of the post-interaction distribution produces an (anomalous) sidebands APINEM spectrum, indicating that the large recoil condition (1) is not a necessary condition for the quantum limit in the QEW regime, and APINEM can be generated in the intermediate regime $2\sigma_t(t_D) > T > \sigma_{t_0}$, (or $2\sigma_z(t_D) > \beta\lambda/2\pi > \sigma_{z_0}$). However, the nature of this APINEM spectrum is quite different from conventional PINEM. The projections of the phase-space displaced branches, overlap in both dimensions, and the APINEM spectrum is a result of their coherent quantum through the mixed first-order term in Eq.5. This produces density modulation in real time and space dimensions with period T ($\beta\lambda$). In the energy (momentum) vertical dimension the horizontal projection of the slanted post-interaction WF, modulated at frequency $\omega$, produces PINEM-like interference fringes of approximate period:

$$\delta E = \frac{\sigma_{E0}}{\sigma_t(t_D)} T, \text{ or } \delta p = \frac{\sigma_{p0}}{\sigma_z(t_D)} \beta\lambda \simeq m^* v_0 \left(\frac{\beta\lambda}{L_D}\right). \qquad (4)$$

We confirm this observation of a universal phase-space classification of electron-wave interactions by explicit solution of the relativistic modified Schrodinger equation [20,17] $i\hbar\partial\psi(z,t)/\partial t = (H_0 + H_I(t))\psi(z,t)$ with the free space Hamiltonian $H_0 = \varepsilon_0 + v_0(-i\hbar\nabla - p_0) + (-i\hbar\nabla - p_0)^2/2m^*$, and with general QEW initial conditions. This is exemplified here for a specific example of SPR (and TR) with an interaction term $H_I(t) = -\frac{e\hbar}{2\gamma_0 m\omega}\left\{e^{-i(\omega t-\phi_0)}\tilde{E}(z)\cdot\nabla - e^{i(\omega t-\phi_0)}\tilde{E}^*(z)\cdot\nabla\right\}$, where $\tilde{E}(z) = \sum_m \tilde{E}_m e^{iq_{zm}z}$ is the near field of a Floquet radiation mode on a grating, interacting with the electron through one of the slow-wave space harmonics $E_0 = \tilde{E}_m, q_z = q_{zm}$ that is synchronous with the electron $v_0 \cong \omega/q_z$. We solved this equation both by first order perturbation theory and by exact numerical computation for the different initial conditions [17].

The third term in the WD together with the first term (zero-order) (3) result in the conventional PINEM spectrum (with only two sidebands in first order perturbation analysis). This corresponds to the quantum plane-wave (PINEM) limit (Fig.1b&c), where $\rho_p^{(f)} = \int \left( W^{(00)}(z,p) + W^{(11)}(z,p) \right) dz \simeq \left( 1 - 2\Upsilon^2 \right) \rho^{(0)}(p) + \Upsilon^2 \left[ \rho^{(0)}(p + \hbar\omega/v_0) + \rho^{(0)}(p - \hbar\omega/v_0) \right]$, and $\rho^{(0)}(p) = |\psi^{(0)}(p)|^2 = (2\pi\sigma_{p_0}^2)^{-\frac{1}{2}} \exp\left( -(p - p_0)^2 / 2\sigma_{p_0}^2 \right)$, $\Upsilon = eE_0 L / 2\hbar\omega$, $E_0$ is the interacting electric component, L is the interaction length.

We draw attention now to the second interference (phase-dependent) term in (3), which is most important in our analysis, because it is the only contribution that can produces the APINEM spectrum effect (Fig.1e), observed in this work, and the 'point-particle-like' acceleration regime[17] (Fig.1d) (see Supp. A): $\rho_p^{(f)} = \int W(z,p,t)\,dz \simeq \rho^{(0)}\left( p - \Delta p^{(1)} \right)$ with momentum shift $\Delta p^{(1)} = \int W(z,p)\,p\,dz = \Delta p_{point}\, e^{-\Gamma^2/2}$, where $\Delta p_{point} = -\frac{eE_0 L}{v_0} sinc\left( \frac{\bar{\theta}}{2} \right) cos\left( \phi_0 + \frac{\bar{\theta}}{2} \right)$ is the classical point-particle acceleration, $\bar{\theta} = (\omega/v_0 - q_z)L$ is the classical "interaction detuning parameter" in FEL theory [3], and $\phi_0$ is the initial relative phase between the electron and the laser-induced field. The decay parameter $\Gamma$ is defined as

$$\Gamma = \left( \frac{\omega}{v_0} \right) \sigma_z(t_D) = \frac{2\pi\sigma_z(t_D)}{\beta\lambda}. \qquad (5)$$

Defining the point-particle-like acceleration regime as the regime where the damping of the QEW acceleration due to the Gaussian decay factor is less than $\Delta p^{(1)} / \Delta p_{point} = 1/e$, then the point-particle regime is: $\Gamma < \sqrt{2}$, or $\sigma_z(t_D) < \beta\lambda/\sqrt{2}\pi$. That is the same (except for factor $\sqrt{2}$) as the "short wavepacket" classical regime condition: the opposite of the "long wavepacket" condition. Using the relation $\sigma_{E_0} \sigma_{t_0} = \hbar/2$, we can now express also the "large recoil" condition (1) for the quantum (PINEM) regime in terms of the factor $\Gamma_0 = 2\pi\sigma_{z_0}/\beta\lambda > \sqrt{2}$, and thus we can define the APINEM regime in the range $\Gamma_0 < \sqrt{2}$ and $\Gamma > \sqrt{2}$: satisfying the "long wavefunction" condition (2), but not the "large recoil" condition (1).

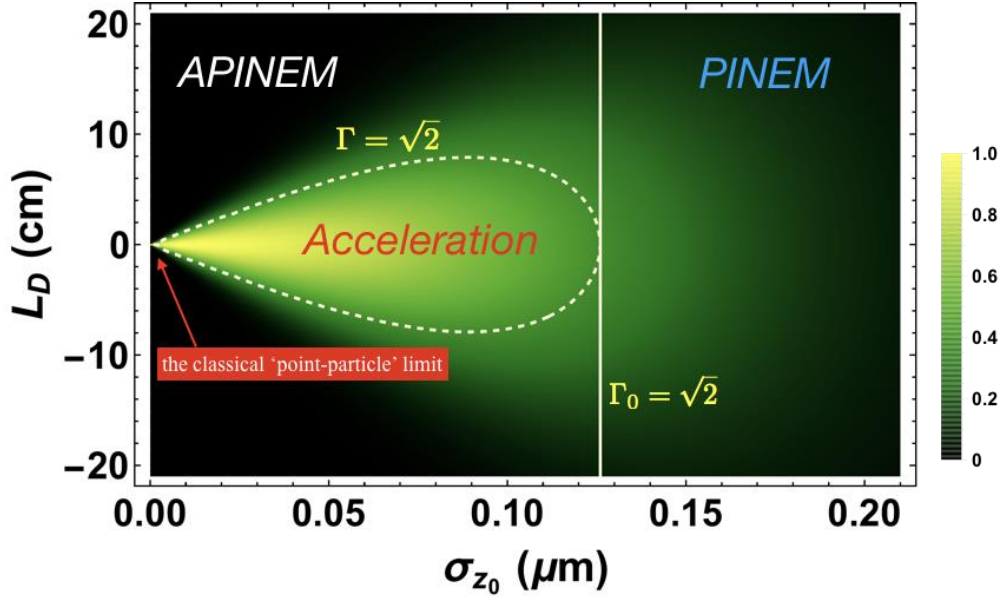

*Fig.2: The classification of PINEM, Acceleration and APINEM regimes in light-matter interaction for optical wavelength* $\lambda = 0.8\,\mu m$. *The classical point particle picture appears at the limit* $\Gamma = 2\pi\sigma_z(L_D)/\beta\lambda \to 0$.

Remarkably, the acceleration decay factor ($e^{-\Gamma^2/2}$) has sole dependence on the QEW size at entrance $\sigma_z(t_D)$ and the /radiation optical wavelength $\beta\lambda$. Considering the dependence of the wavepacket size $\sigma_z(t_D) = \sqrt{\sigma_{z_0}^2 + \left(\dfrac{\lambda_c^*}{4\pi}\dfrac{ct_D}{\sigma_{z_0}}\right)^2}$ on its minimal spot-size $\sigma_{z_0}$ and its pre-interaction drift length $L_D = v_0 t_D$, it is instructive to display the Gaussian factor as a function of these two parameters. In Fig.2 we display its dependence on ($\sigma_{z_0}, L_D$) in color code for the particular parameters example $\beta = v_0/c = 0.7$, $\lambda = 0.8\,\mu m$ [11]. Note that the diagram includes negative drift length $L_D < 0$, corresponding to the case of a QEW entering the interaction region with a converging phase and negative chirp.

The dashed contour in the phase diagram (Fig.2) marks the transition border $\Gamma(\sigma_{z_0}, L_D) = \sqrt{2}$, within which the QEW exhibits 'point-particle-like' acceleration (the "acceleration" regime) corresponding to Fig.1d. The vertical solid line defines the transition point $\Gamma_0 = \sqrt{2}$, beyond which the QEW displays discrete PINEM-kind sideband energy spectrum (the "PINEM" regime –

Fig.1b&c). The third zone $\Gamma_0 < \sqrt{2} < \Gamma$ defines the APINEM regime (Fig. 1e). Only in the small recoil regime $\Gamma_0 < \sqrt{2}$ it is possible to demonstrate the transition from the point-particle-like acceleration regime to the quantum APINEM interference regime by performing the radiative interaction after different drift lengths $L_D$, crossing the curve $\Gamma_0 = \sqrt{2}$. For given $\sigma_{z_0}$, transition from the acceleration to PINEM regime can be demonstrated by changing the interaction wavelength.

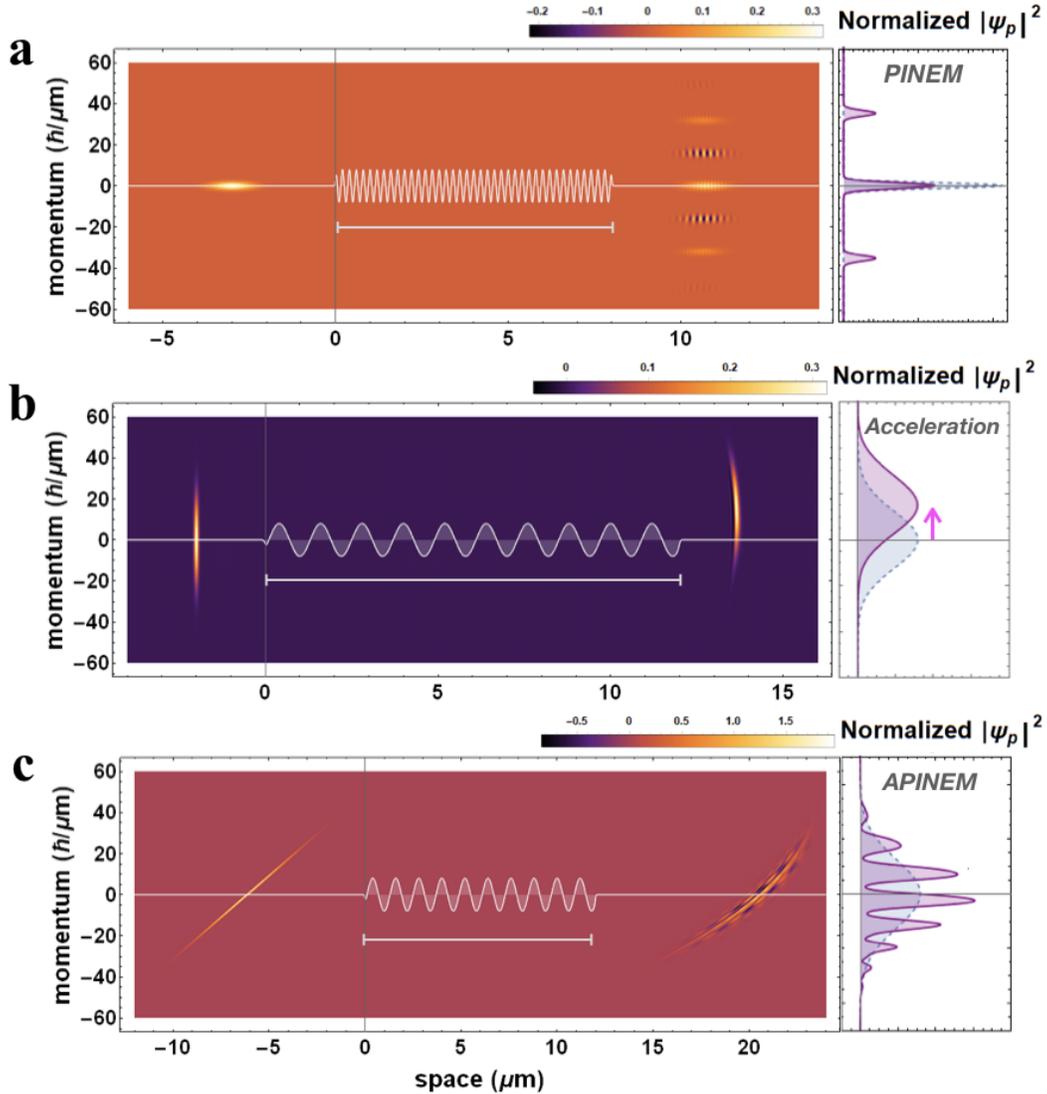

*Fig.3: Numerical simulations of (a) PINEM, (b) Acceleration and (c) APINEM for a quantum electron wavepacket passing through the near-field of a grating, illuminated by a laser beam. Shown are the evolving quantum/classical features in phase space and the momentum distributions in momentum space.*

*Simulation Setup* - To confirm our observations beyond perturbation analysis, we demonstrate the quantum characteristics of electron-light interaction in phase-space through numerical solution (see sup. B) of the Schrodinger equation for the example of stimulated SPR interaction of a QEW with the near field of a grating that we have analyzed earlier with semiclassical [17] and quantum electrodynamics formulations [16].

Figure 3 displays the numerically computed phase-space evolution of a single QEW for different initial conditions. Fig.3a shows quantum regime sideband momentum spectrum in a wave-like interaction regime for parameters of a grating period $\lambda_G = 0.2\,\mu m$ (with synchronizm condition satisfied), intrinsic wavepacket size $\sigma_{t_0} = 1.9\,\text{fts}$ ($\sigma_{z_0} = 0.4\,\mu m$), $\beta = v_0/c = 0.7$. Fig.3b shows the point-particle-like classical acceleration in the short wavepacket (small recoil) regime for $\beta_0\lambda = 1.2\,\mu m$, $\sigma_{z_0} = 0.04\,\mu m$ ($\sigma_{t_0} = 0.2\,\text{fts}$) and relative phase $\phi_0 = 0$, consistent with [17]. The new APINEM case is demonstrated in Fig.3c for the same parameters of Fig.3b, but with a wavepacket entering stretched and chirped after a pre-drift length of $L_D = 60\,\text{cm}$, such that the long wavepacket condition $\sigma_z(L_D) = 1.5\,\mu m > \beta_0\lambda$ is satisfied. It thus displays the emergence of the quantum interference branches in the WD and the interference fringes in the momentum distribution.

Figure 4 shows the dependence of the APINEM fringes period on the incident reduced radiation wavelength $\beta\lambda$, calculated for the two sets of parameters $\sigma_{z_0} = 0.04\,\mu m$, $L_D = 40\,\text{cm}$ (red points), and $\sigma_{z_0} = 0.06\,\mu m$, $L_D = 60\,\text{cm}$ (purple points), well matching the analytical expression (4) (two straight lines). Note that the linear dependence of the APINEM fringes period on the radiation wavelength is drastically different (inverse) than the sidebands period of the PINEM spectrum (dashed line, and blue points).

It is necessary to stress the significance of the quantum interference fringes in producing the PINEM and APINEM spectra. In the APINEM case (1d), the multi-sidebands spectrum is generated due to the horizontal interference of fringes that are generated by the mixed interference term in (3) ($2\Re(W^{(01)}) \propto \cos(\omega(t - z/v_0))$). This presently reported new term, is therefore essential for identifying the APINEM interaction regime (1e) and the transition from quantum sidebands (1d) to classical acceleration (1c) in the QEW regime. This coherent phase space-dynamics appears to be similar to the interference features of Schrodinger's cat states with superposition of two coherent states in quadrature phase-space of quantum light. [15]

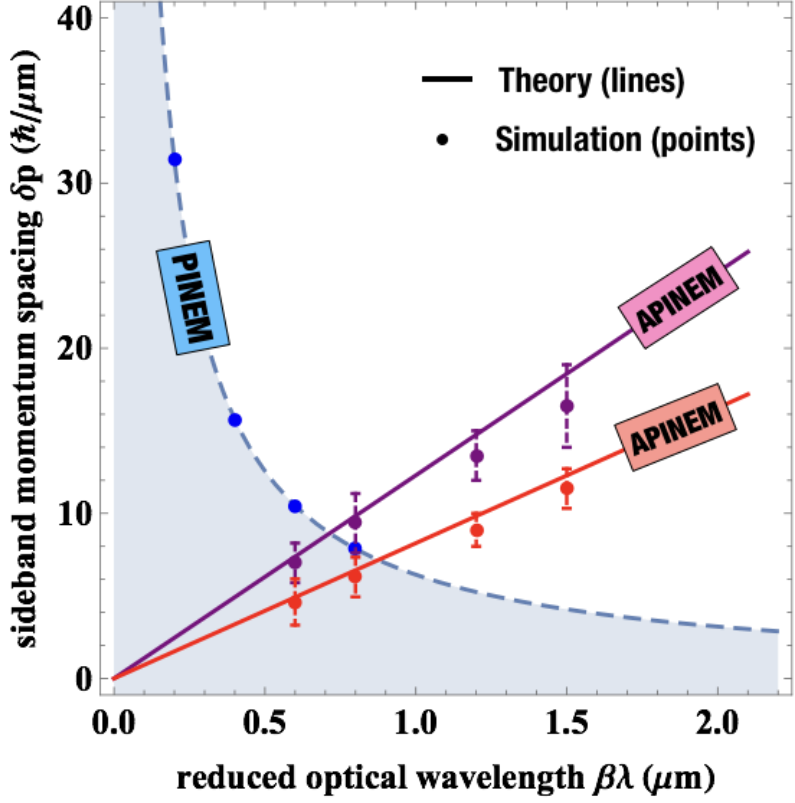

*Fig.4 The momentum/energy spectral sideband spacing in PINEM ( $\delta p = \hbar\omega/v_0 = 2\pi\hbar/\beta\lambda$ ) and APINEM (Eq.4). The lines (solid and dashed) are from theory, and correspondingly the points (blue, red and purple) from simulations for different wavepacket sizes and drift lengths.*

*Measurement Limits* - For practical measurement of the energy spectrum of individual electrons, an ensemble of particles or interaction event measurements has to be accounted statistically in an energy spectrum analyzer. This requires averaging of the single electron spectra over a classical statistical distribution of the ensemble that depends on the electron source (cathode temperature, gun voltage stability etc.) [17,21], and thus adds an extra classical uncertainty term ( $\sigma_{part}$ ) into the electron beam ensemble spread $\Delta E = 2\sqrt{\sigma_{E_0}^2 + \sigma_{E,part}^2}$ . While in the single QEW regime, the APINEM quantum interference regime (Fig.2) exists for $\sigma_{E0} > \hbar\omega$ , in violation of the large recoil condition (1), it is evident that the interference pattern would wash out in an ensemble, unless $\sigma_{E,part} < \hbar\omega$ . Therefore, the ensemble should satisfy $\sigma_{E,part} < \sigma_{E0}$ . Since in conventional cathode electron guns (perhaps with the exception of some exotic electron sources [22]), the situation is opposite (the classical ensemble energy uncertainty is dominant), a preselection phase-space filtering process is required to build up the required ensemble. Furthermore, both the APNEM and particle acceleration are phase sensitive, and require phase-locking of the electron entrance time to the interaction region relative to the laser $\Delta\phi_0 = \omega\Delta t_0 < 2\pi, (\Delta t_0 < T)$. Thus, the transition from the acceleration regime to APINEM and PINEM regimes, can be demonstrated experimentally only with a properly preselected ensemble of electrons in phase-space. Such may possibly be realizable with advancement of single electron wavepacket phase-space control and filtering based on optical (or THz) streaking techniques [18-19].

*Conclusion* - In this work, we have revealed the different classical and quantum interaction regimes of free electrons with radiation. These are uniquely determined by the topology of the coherent (minimal Heisenberg uncertainty) QEW in phase-space representation: point-particle-like acceleration regime, near-field photo-induced interactions (PINEM) regime, and a newly reported anomalous PINEM regime. The model demonstrates the transition from the classical point-particle to quantum wavefunction interaction regime, thus resolving the particle-wave duality question in the context of radiative interactions, and assigning measurable physical reality to the history-dependent dimensions of the interacting electron wavefunction.

## Acknowlegements

The work was supported in parts by DIP (German-Israeli Project Cooperation), The Israel Science Foundation and by the PBC program of the Israel council of higher education. Correspondance and requests for materials should be addressed to A. G. (gover@eng.tau.ac.il) or Y. P. (yiming.pan@weizmann.ac.il).

## References:

1. Madey, J. M. (1971). Stimulated emission of bremsstrahlung in a periodic magnetic field. *Journal of Applied Physics*, *42*(5), 1906-1913.
2. Pellegrini, C., Marinelli, A., & Reiche, S. (2016). The physics of x-ray free-electron lasers. *Reviews of Modern Physics*, *88*(1), 015006.
3. Gover, A., and P. Sprangle, *IEEE Journal of Quantum Electronics* 17(7), 1196-1215 (1981).

4. Peralta, E. A., Soong, K., England, R. J., Colby, E. R., Wu, Z., Montazeri, B., & Sozer, E. B. (2013). Demonstration of electron acceleration in a laser-driven dielectric microstructure. *Nature*, *503*(7474), 91-94.
5. Breuer, J., & Hommelhoff, P. (2013). Laser-based acceleration of nonrelativistic electrons at a dielectric structure. *Physical review letters*, 111(13), 134803.
6. Smith, S. J., & Purcell, E. M. (1953). Visible light from localized surface charges moving across a grating. *Physical Review*, *92*(4), 1069.
7. Cherenkov, P. A., *Doklady Akademii Nauk SSSR.* 2, 451(1934).
8. V. L. Ginzburg and I. M. Frank, *Zh. Eksp. Teor. Fiz.* 16, 15–22 (1946).
9. Barwick, B., Flannigan, D. J., & Zewail, A. H. (2009). Photon-induced near-field electron microscopy. *Nature*, *462*(7275), 902-906.
10. Garcia de Abajo, F. J., Asenjo-Garcia, A., Kociak, M. Multiphoton absorption and emission by interaction of swift electrons with evanescent light _elds. Nano letters, 10(5), 1859-1863 (2010).
11. Feist, A., Echternkamp, K. E., Schauss, J., Yalunin, S. V., Schfer, S., Ropers, C. Quantum coherent optical phase modulation in an ultrafast transmission electron microscope. Nature, 521(7551), 200-203 (2015).
12. Priebe, K. E., Rathje, C., Yalunin, S. V., Hohage, T., Feist, A., Schäfer, S., & Ropers, C. (2017). Attosecond electron pulse trains and quantum state reconstruction in ultrafast transmission electron microscopy. *Nature* Photonics, 11(12), 793.
13. Echternkamp, K. E., Feist, A., Schäfer, S., & Ropers, C. (2016). Ramsey-type phase control of free-electron beams. Nature Physics, 12(11), 1000.
14. Kling, P., Giese, E., Endrich, R., Preiss, P., Sauerbrey, R., & Schleich, W. P. (2015). What defines the quantum regime of the free-electron laser? *New Journal of Physics*, *17*(12), 123019.
15. Schleich, Wolfgang P. *Quantum optics in phase space*. John Wiley & Sons, 2011.
16. Yiming Pan, Avraham Gover, Spontaneous and Stimulated Emissions of Quantum Free-Electron Wavepackets-QED Analysis. arXiv preprint arXiv:1805.08210 (2018).
17. Gover, Avraham, and Pan, Yiming. Dimension-dependent stimulated radiative interaction of a single electron quantum wavepacket. Physics Letters A 382(23), 1550-1555 (2018).
18. Baum, P. (2017). Quantum dynamics of attosecond electron pulse compression. Journal of Applied Physics, 122(22), 223105.
19. Kealhofer, C., Schneider, W., Ehberger, D., Ryabov, A., Krausz, F., & Baum, P. (2016). All-optical control and metrology of electron pulses. Science, 352(6284), 429-433.

20. Friedman, A., Gover, A., Kurizki, G., Ruschin, S., & Yariv, A., *Reviews of Modern Physics* **60(2)**, 471 (1988).

21. Ford, G. W., and R. F. O'connell. *American Journal of Physics* **70.3**: 319-324(2002).

22. Franssen, J. G. H., Frankort, T. L. I., Vredenbregt, E. J. D., & Luiten, O. J. (2017). Pulse length of ultracold electron bunches extracted from a laser cooled gas. Structural Dynamics, 4(4), 044010.